\documentclass[twocolumn]{jpsj3}
\usepackage{graphicx}
\usepackage{amsmath,amssymb}

\title{Locally Non-centrosymmetric Superconductivity in Multilayer Systems}

\author{Daisuke MARUYAMA\thanks{E-mail address: marudai@phys.sc.niigata-u.ac.jp}, Manfred SIGRIST$^{1}$, and Youichi YANASE$^{2}$ 
}
\inst{
Graduate School of Science and Technology, Niigata University, Niigata 950-2181, Japan \\ $^{1}$Theoretische Physik, ETH-Honggerberg, 8093 Zurich, Switzerland \\ $^{2}$Department of Physics, Niigata University, Niigata
950-2181, Japan 
}

\abst{Although multilayer systems possess {\it global} inversion symmetry, 
some of the layers lack {\it local} inversion symmetry because  
no global inversion centers are present on such layers. 
Such locally non-centrosymmetric systems exhibit spatially modulated 
Rashba spin-orbit coupling. 
In this study, the superconductivity in multilayer models exhibiting inhomogeneous 
Rashba spin-orbit coupling is investigated. 
We study the electronic structure, superconducting gap, and spin 
susceptibility in the superconducting state with mixed parity 
order parameters. 
We show the enhancement of the spin susceptibility by Rashba spin-orbit 
coupling and interpret it on the basis of the crossover from 
a centrosymmetric superconductor to a non-centrosymmetric 
superconductor. 
It is also shown that the spin susceptibility is determined 
by the phase difference of the order parameter between layers and is nearly 
independent of the parity mixing of order parameters. 
An intuitive understanding is given on the basis of the analytic expression 
of superconducting order parameters in the band basis. 
The results indicate that not only a broken {\it global} inversion 
symmetry but also a broken {\it local} inversion symmetry leads to 
unique properties of superconductivity. 
We discuss the superconductivity in artificial superlattices 
involving CeCoIn$_5$ and multilayer high-$T_{\rm c}$ cuprates.  
}

\kword{superconductivity without local inversion symmetry, spin susceptibility, multilayer system}

\begin{document}
\maketitle

\newcommand{\etal}{{\it et al.}: }
\newcommand{\Pt}{CePt$_3$Si }
\newcommand{\Rh}{CeRhSi$_3$ }
\newcommand{\Ir}{CeIrSi$_3$ }
\newcommand{\Ptf}{CePt$_3$Si}
\newcommand{\Rhf}{CeRhSi$_3$}
\newcommand{\Irf}{CeIrSi$_3$}
\newcommand{\PRL}{Phys. Rev. Lett. } 
\newcommand{\PRB}{Phys. Rev. B } 
\newcommand{\JPSJ}{J. Phys. Soc. Jpn. } 
\newcommand{\kx}{k_{\rm x}}
\newcommand{\ky}{k_{\rm y}}
\renewcommand{\k}{\boldsymbol{k}}

\section{Introduction}

 The discovery of superconductivity in the 
heavy-fermion compound without inversion symmetry, CePt$_3$Si, 
\cite{PhysRevLett.92.027003} triggered 
extensive studies of non-centrosymmetric superconductivity~\cite{Springer}. 
Subsequently, several new non-centrosymmetric superconductors 
with unique properties have been identified among heavy-fermion materials and in other 
classes of materials~\cite{
LowTempPhys.31.748,JPSJ.76.051009,PhysRevLett.95.247004,JPSJ.76.051010, 
JPSJ.75.043703,JPSJ.76.051003,rf:CeCoGe,JPSJ.73.3129,rf:togano,
rf:akimitsu,rf:shibayama,rf:mu,rf:klimczuk,rf:zuev,rf:kreiner}.
The research field has even been effectively extended to 
non-centrosymmetric superfluids in cold Fermi gases
~\cite{Sato-Fujimoto,Iskin} in which  
antisymmetric spin-orbit coupling is artificially induced. 
Many interesting properties, such as the parity mixing of order 
parameters~\cite{Sov.Phys.68.1244}, 
the magnetoelectric effect \cite{PhysRevB.72.172501,
PhysRevB.65.144508,PhysRevB.72.024515,JPSJ.76.034712},
anisotropic spin susceptibility \cite{Sov.Phys.68.1244,
Sov.Phys.44.1243,PhysRevLett.87.037004,
PhysRevLett.92.097001,NewJPhys.6.115,PhysRevLett.94.027004,
PhysRevB.72.212504,JPSJ.76.043712,JPSJ.77.124711}
accompanied by an anomalous paramagnetic depairing
effect\cite{JPSJ.76.124709}, 
helical superconducting phases in a magnetic field
\cite{JPSJ.76.124709,JETP.Lett.78.637,
PhysRevB.76.014522,PhysRevB.70.104521,PhysRevLett.94.137002,
PhysRevB.75.064511,Matsunaga-Ikeda}, 
and topological superfluid phases~\cite{Sato-Fujimoto}, 
have been proposed and studied in various contexts. 

\begin{figure}[htbp]
\begin{center}
\includegraphics[scale=0.4]{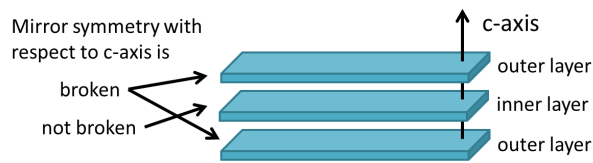}
\end{center}
\caption{(Color online) Schematic view of trilayer system with focus 
on local inversion symmetry. 
The inversion center is present in the inner layer; it is absent 
in the outer layers. This crystal structure is regarded as a 
locally non-centrosymmetric system. 
}
\label{schematic}
\end{figure}

The term non-centrosymmetric superconductivity is commonly used for
systems that have no inversion center. It is, however, interesting
to note that features familiar in non-centrosymmetric superconductors
can also be relevant for systems with an inversion center, but specific forms
of a local violation of inversion symmetry. 
A clear example is a multilayer system in which the 
local mirror symmetry is broken, as shown in Fig.~\ref{schematic}. 
Naturally, the question arises whether superconductivity displays 
any exotic property in such a {\it locally} non-centrosymmetric system. 
One of the authors has earlier investigated this issue by studying the 
spin triplet superconducting state with random Rashba spin-orbit 
coupling arising from stacking faults~\cite{Yanase_randomtriplet}. 
In this study, we focus on regular multilayer systems and elucidate
the basic properties of such {\it locally} non-centrosymmetric 
superconductors. We show that, indeed, some properties of the superconducting phase
can be strongly affected by a broken local inversion symmetry in multilayer 
systems that exhibit small interlayer coupling.

 Our study is motivated by the observation of superconductivity in  
artificial superlattices consisting of the heavy-fermion superconductor CeCoIn$_5$ 
and the conventional metal YbCoIn$_5$ \cite{Science.327.980,private.superlattices}. 
Although it is expected that the superconductivity occurs in sufficiently thick multilayer 
structures of CeCoIn$_5$, it is surprising that it even prevails down to superlattice incorporating 
stacks of only three well-separated CeCoIn$_5$ layers.

 The superconductivity in bulk CeCoIn$_5$ has attracted considerable 
interest for its unique magnetic features, such as the paramagnetic depairing 
effect,~\cite{Tayama} the possible realization of the Fulde-Ferrell-Larkin-Ovchinnikov phase 
at high magnetic fields,~\cite{radovan2003,PhysRevLett.91.187004}
an unconventional magnetic order,~\cite{young2007,kenzelmann2008}
and field-induced quantum criticality.~\cite{paglione2003,bianchi2003,ronning2005,izawa2007,panarin2009}  
These features make it  even more attractive to investigate the superconductivity in 
superlattices of CeCoIn$_5$. As such, it represents an ideal 
system for studying the effects of {\it local} inversion 
symmetry breaking, and more so because many of the striking properties of non-centrosymmetric 
superconductors may result from the additional presence of strong magnetic correlations.  

A better class of systems of similar character is that of multilayer high-$T_{\rm c}$
cuprates\cite{Mukuda,Shimizu-Mukuda,PhysRevLett.94.137003}. 
Although cuprate superconductors have been intensively investigated 
since their discovery in 1986, the role of spatially modulated Rashba 
spin-orbit coupling has not received much attention thus far.  
Since the magnetic properties of multilayer cuprates are 
investigated by nuclear magnetic resonance (NMR) measurement, 
it is also possible to study multilayer cuprates focusing on 
local non-centrosymmetricity. 
For instance, the spin susceptibility in the 
superconducting state can be measured with a layer resolution
using the Knight shift of NMR. 
Following a brief report on the spin susceptibility in locally 
non-centrosymmetric systems~\cite{Maruyama_LT}, 
we study here the electronic structure, 
superconducting gap, and spin susceptibility of multilayer 
superconductors in detail. 

The paper is organized as follows. In \S2, we introduce the model 
Hamiltonian for multilayer systems including spatially modulated 
Rashba spin-orbit coupling and parity-mixed superconducting order parameters. 
In \S3, we show the numerical results of the spin susceptibility 
in the bi- and tri-layer systems to demonstrate 
the crossover from a conventional superconductor to a 
non-centrosymmetric superconductor. 
The numerical results are discussed on the basis of analytic 
expressions for the electronic structure (\S4) and 
superconducting gap (\S5) by decomposing the spin susceptibility 
into the Pauli and Van-Vlecks parts (\S6). 
The relationship between the symmetry of order parameters and 
the spin susceptibility is clarified in \S7, where a nontrivial role of 
the interlayer phase difference of order parameters is discussed. 
The numerical results of the spin susceptibility in more than three layers 
are shown in \S8. 
A brief summary and discussion of the superconductivity in CeCoIn$_5$ 
are given in \S9.

\section{Model}

First, we introduce a model Hamiltonian for two-dimensional 
multilayer superconductors with spatially modulated Rashba 
spin-orbit coupling as 
{\setlength\arraycolsep{1pt}
\begin{eqnarray}
\label{model}
H&=&H_{\rm b}+H_{\rm SOC}+H_{\rm pair}+H_{\perp},
\\
H_{\rm b}&=&
\sum_{\mib{k},s,m} \varepsilon(\mib{k}) \, c^{\dag}_{\mib{k}sm}c_{\mib{k}sm},
\\ 
H_{\rm SOC}&=&
\sum_{\mib{k},s,s',m} \alpha_{m} \, 
\mib{g}(\mib{k})\cdot\mib{\sigma}_{ss'}c^{\dag}_{\mib{k}sm}c_{\mib{k}s'm},
\\ 
H_{\rm pair}&=&
\frac{1}{2}\sum_{\mib{k},s,s',m}[\Delta_{ss'm}(\mib{k}) \, 
c^{\dag}_{\mib{k}sm}c^{\dag}_{-\mib{k}s'm}+\rm{h.c.}], \\
H_{\perp}&=&t_{\perp}\sum_{\mib{k},s,\langle m,m'\rangle}
c^{\dag}_{\mib{k}sm}c_{\mib{k}sm'},
\end{eqnarray}}
where $c_{\mib{k}sm}$ ($c^{\dag}_{\mib{k}sm}$) is the annihilation
(creation) operator for an electron with spin $s$ on layer $m$, and
$\mib{\sigma}_{ss'}$ is the vector representation of the Pauli matrix. 

The first term, $H_{\rm b}$, describes the dispersion relation without
spin-orbit coupling or interlayer coupling. 
We consider a square lattice and assume a tight-binding model,
i.e., $\varepsilon(\mib{k})=-2t(\cos{k_{x}}+\cos{k_{y}})-\mu$. 
The $(x,y,z)$-axes correspond to the $(a,b,c)$-axes of the tetragonal
crystal structure. We choose the unit of energy as $t=1$ and fix the chemical potential
$\mu=-1$, which leads to the electron density per site being approximately 0.63. The
following results are nearly independent of the dispersion relation 
and electron density. 

The second term, $H_{\rm SOC}$, describes the Rashba spin-orbit coupling 
arising from the lack of local inversion symmetry. 
This term preserves time reversal symmetry, if the g-vector is odd 
in $\mib{k}$, i.e., $\mib{g}(-\mib{k})=-\mib{g}(\mib{k})$. 
The coupling constants $\alpha_{m}$ should have opposite signs 
between layers above and below the inversion center so as to 
conserve global inversion symmetry. 
For instance, $(\alpha_1, \alpha_2, \alpha_3) = (\alpha, 0, -\alpha)$ 
for trilayers. 
For $\mib{g}(\mib{k})$, the multilayer systems should have a 
Rashba-type g-vector 
because the mirror symmetry is broken for outer layers 
(see Fig.~\ref{schematic}). 
Although the detailed momentum dependence of $\mib{g}(\mib{k})$ 
is determined by electronic structures \cite{JPSJ.77.124711}, 
we assume here the simple form 
$\mib{g}(\mib{k})=(-\sin{k_{y}},\sin{k_{x}},0)$. 

The third term, $H_{\rm pair}$, introduces intralayer Cooper pairing 
via an off-diagonal mean field. We ignore interlayer 
pairing as we consider the layers to be weakly coupled. Since we take into account the spatially 
modulated Rashba spin-orbit coupling arising from a broken local
inversion symmetry, the order parameter $\Delta_{ss'm}({\bf k})$ involves 
both spin singlet and triplet components, 
{\setlength\arraycolsep{1pt}
\begin{eqnarray}
\Delta_{ss'm}(\mib{k}) 
=\left(
\begin{array}{ccc}
-d_{{\rm x}m}(\mib{k})+{\rm{i}}d_{{\rm y}m}(\mib{k}) & 
\psi_{m}(\mib{k})+d_{{\rm z}m}(\mib{k}) \\
-\psi_{m}(\mib{k})+d_{{\rm z}m}(\mib{k}) & 
d_{{\rm x}m}(\mib{k})+{\rm{i}}d_{{\rm y}m}(\mib{k})
\label{op}
\end{array}
\right), \nonumber \\
\end{eqnarray}}
where $\psi_{m}(\mib{k})$ and $\mib{d}_{m}(\mib{k})$ are the corresponding scalar and
vector order parameters for the spin singlet and triplet pairings on the
layer $m$, respectively. For our discussion, we avoid the use of a
microscopic model based on a pairing mechanism. Rather, we introduce
an order parameter on phenomenological grounds, which we assume to have
$s$-wave symmetry for the
singlet pairing and a $p$-wave symmetry for the triplet pairing. On
symmetry grounds, we choose 
$\psi_{m}(\mib{k})=\psi_{m}$ and
$\mib{d}_{m}(\mib{k})=d_{m}\mib{g}(\mib{k})=d_{m}(-\sin{k_{y}},\sin{k_{x}},0)$. We
take $|\psi_{m}|$, $|d_{m}| \leq 0.01$ to be sufficiently small to satisfy the
condition $|\Delta_{ss'm}(\mib{k})|\ll|\alpha_{m}|\ll\varepsilon_{\rm
F}$ ($\varepsilon_{\rm F}$ is the Fermi energy), as realized in most
non-centrosymmetric superconductors. To minimize the interlayer coupling energy, 
the dominant order parameter component maintains the same sign over all layers, 
whereas the other (subdominant) component adjusts the sign to that of spin-orbit 
coupling ($\alpha_{m}$).

The fourth term, $H_{\perp}$, describes the interlayer hopping of electrons between nearest-neighbor layers. Since we consider a quasi-two-dimensional system, we assume that the interlayer hopping $t_{\perp}$ is smaller than the intralayer hopping $t$.

\section{Numerical Results of Spin Susceptibility}

\subsection{Spin susceptibility for fields along $c$-axis}

We now determine the spin susceptibility of multilayer 
superconductors for spatially modulated Rashba spin-orbit coupling, 
with a magnetic field applied along the $c$-axis.
The spin susceptibility 
$\chi={\rm lim}_{H \rightarrow 0} \langle M_s \rangle/ {H}$ 
is obtained by calculating the magnetization $\langle M_s \rangle$ 
in the field $\mib{H}$.  
The Zeeman coupling term 
is introduced as
\begin{eqnarray}
H_{\rm Z}=- \frac{g \mu_{\rm B}}{2} \sum_{\mib{k},s,s',m}\mib{H}\cdot\mib{\sigma}_{ss'}c^{\dag}_{\mib{k}sm}c_{\mib{k}s'm}, 
\end{eqnarray}
where we assume $g=2$ and $\mu_{\rm B}$ is the Bohr magneton. 
First, the Hamiltonian is diagonalized by introducing
the unitary transformation $ \hat{C}^{\dag}_{\mib{k}} = \hat{\Gamma}^{\dag}_{\mib{k}} \hat{U}^{\dag}(\mib{k})$
in the Nambu space of $ M $ layers, where quasi-particle operators form a $4M$-dimensional vector,
{\setlength\arraycolsep{1pt}
\begin{eqnarray}
\hat{C}^{\dag}_{\mib{k}}=(c^{\dag}_{\mib{k} \uparrow 1},&& c^{\dag}_{\mib{k} \downarrow 1}, c_{-\mib{k} \uparrow 1}, c_{-\mib{k} \downarrow 1}, \dots, \nonumber \\  &&,\dots, c^{\dag}_{\mib{k} \uparrow M}, c^{\dag}_{\mib{k} \downarrow M}, c_{-\mib{k} \uparrow M}, c_{-\mib{k} \downarrow M}),
\end{eqnarray}}
and are analogous to the Bogoliubov quasi-particle operators $
\hat{\Gamma}^{\dag}_{\mib{k}} = (\gamma^{\dag}_{\mib{k} 1},
\gamma^{\dag}_{\mib{k} 2 }, \dots ,\gamma^{\dag} _{\mib{k}  4M})
$. The diagonalized Hamiltonian is written as 
\begin{eqnarray}
H + H_{\rm Z} = \frac{1}{2}\sum_{\mib{k}}\sum_{i=1}^{4M} E_i(\mib{k}) \gamma^{\dag}_{\mib{k} i} \gamma_{\mib{k} i} ,
\end{eqnarray}
where $ E_i(\mib{k}) $ is the quasi-particle energy. 
The magnetization is given as
\begin{eqnarray}
\langle
 M_s \rangle=  \frac{g\mu_{\rm B}}{2}\sum_{\mib{k}}\sum^{4M}_{i=1}[\hat{S}^z (\mib{k})]_{ii}f(E_{i}(\mib{k})),
\end{eqnarray}
where $ f(E) $ is the Fermi-Dirac distribution function. 
The matrix representation of a spin operator is defined on the 
$\hat{\Gamma}^{\dag}_{\mib{k}}$ basis as
\begin{eqnarray}
\hat{S}^{\mu}(\mib{k})=\hat{U}^{\dag}(\mib{k}) \hat{\Sigma}^{\mu} \hat{U}(\mib{k}),
\end{eqnarray}
with $\hat{\Sigma}^{\mu}$ as the $\mu$-component of the spin operator 
in the $4M$-dimensional space. 

We now focus on the spin susceptibility of bilayer ($M=2$) and
trilayer ($M=3$) systems at zero temperature. More than three layers 
will be discussed in \S8. 
The corresponding coupling constants of Rashba spin-orbit coupling 
are described as $(\alpha_1, \alpha_2) = (\alpha, -\alpha)$ for bilayers and $(\alpha_1, \alpha_2, \alpha_3) = (\alpha, 0, -\alpha)$ for trilayers.

\begin{figure}[htbp]
\begin{center}
\includegraphics[scale=0.35]{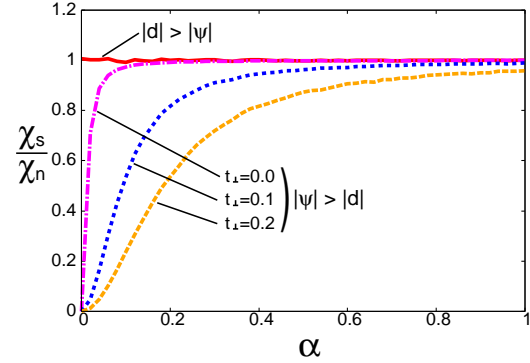}
\end{center}
\caption{(Color online) 
Spin susceptibility along $c$-axis for bilayer system.
The interlayer coupling $t_\perp$ is shown in the figure for the 
dominantly spin singlet pairing state 
(dashed, dotted, and dash-dotted lines). The spin susceptibility is 
independent of interlayer coupling when the spin triplet channel 
is dominant (solid line).  
We assume that $\psi=0.01$ and $d=0$ in the former, and $\psi=0$ and $d=0.01$ 
in the latter. 
The spin susceptibility is nearly independent of the subdominant 
order parameter, as will be shown in Fig.~\ref{st2}.
}
\label{2lay}
\end{figure}

\begin{figure}[htbp]
\begin{center}
\includegraphics[scale=0.35]{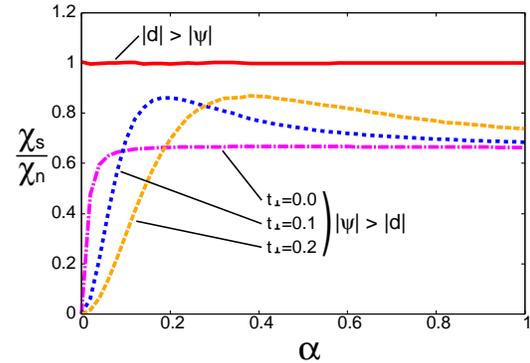}
\end{center}
\caption{(Color online) 
Spin susceptibility along $c$-axis for trilayer system.
The interlayer coupling $t_\perp$ is shown in the figure for the 
dominantly spin singlet pairing state 
(dashed, dotted, and dash-dotted lines). The spin susceptibility is 
independent of interlayer coupling when the spin triplet channel 
is dominant (solid line).  
The spin susceptibility is nearly independent of the subdominant 
order parameter as in the bilayer system (Fig.~\ref{st2}). 
}
\label{3lay}
\end{figure}

We compare two cases: (A) the spin triplet channel is dominant
$|d|>|\psi|$ and (B) the spin singlet channel is dominant
$|\psi|>|d|$. The order parameters are assumed as follows:
\begin{description}
\item[({\rm A})]
$(\psi_{1}, \psi_{2}) = (\psi, -\psi)$ and $(d_{1}, d_{2}) = (d, d)$ 
for bilayers. $(\psi_{1}, \psi_{2}, \psi_{3}) = (\psi, 0, -\psi)$ and 
$(d_{1}, d_{2}, d_{3}) = (d, d, d)$ for trilayers.
\item[({\rm B})]
$(\psi_{1}, \psi_{2}) = (\psi, \psi)$ and $(d_{1}, d_{2}) = (d, -d)$ 
for bilayers. $(\psi_{1}, \psi_{2}, \psi_{3}) = (\psi, \psi, \psi)$ and 
$(d_{1}, d_{2}, d_{3}) = (d, 0, -d)$ for trilayers.
\end{description}
In case (A), the spin susceptibility remains 
unaffected by the superconductivity $\chi_{\rm s} = \chi_{\rm n}$
because the spin triplet component of the type $
\mib{d}_m(\mib{k})\propto\mib{g}(\mib{k})\perp\hat{z}$ is an equal-spin
pairing state with Cooper pair spins along the $c$-axis. Thus, spin
polarization in the superconducting phase is possible without pair
breaking. This feature is essentially independent of spin-orbit 
coupling, interlayer hopping, and the number of layers, as can be seen in Figs.~\ref{2lay} and \ref{3lay} for both bi- and tri-layer systems, respectively. 
We will provide more rigorous arguments for this kind of behavior in \S7. 

\begin{figure}[htbp]
\begin{center}
\includegraphics[scale=0.35]{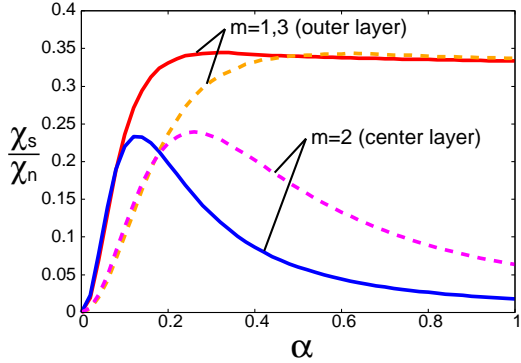}
\end{center}
\caption{(Color online) 
Contribution of each layer to spin susceptibility 
along $c$-axis in trilayer system. 
We assume the spin singlet state ($\psi=0.01$ and $d=0$)  
with $t_{\perp}=0.1$ (solid line) and $t_{\perp}=0.2$ (dashed line).}
\label{3bunlay}
\end{figure}

More interesting is case (B), as spin singlet pairing leads to
a complete suppression of the spin susceptibility at $T=0$ in a
conventional superconductor. Indeed for a vanishing 
spin-orbit coupling ($\alpha=0$), we
find $\chi_{\rm s}=0$ irrespective of $t_{\perp}$. As soon as 
spin-orbit coupling is turned on, however, the spin susceptibility 
gradually recovers and
approaches a constant value for large $\alpha$: $\chi_{\rm
s}\to\chi_{\rm n}$ for the bilayer and $\chi_{\rm s}\to 2\chi_{\rm
n}/3$ for the trilayer. The mechanism of this behavior lies in the
band structure affected by Rashba-type spin-orbit coupling (see \S4).
Note that the layers behave as being nearly decoupled for large $\alpha$, 
and then the spin susceptibility along $c$-axis is recovered 
for layers with non-vanishing $\alpha_m$. 
Consequently, in the bilayer system, all layers are involved,
giving rise to a full recovery of $\chi_{\rm s}$ for large $\alpha$
(analogous to the uniformly non-centrosymmetric superconductor
\cite{NewJPhys.6.115}), whereas in the trilayer system, only two of the three
layers can contribute to what yielding a correspondingly reduced limiting value of
$\chi_{\rm s} = 2\chi_{\rm n}/3$. 
Figure~\ref{3bunlay} corroborates this picture by
considering the contributions of the different layers. Indeed, in a
large-$\alpha$ regime, the outer layers $m=1,3$ carrying spin-orbit coupling saturate at
$\chi_{\rm s}\to \chi_{\rm n}/3$, while the center layer $m=2$ completely
suppresses the spin susceptibility. 
Remarkably, at small $\alpha$ ($<t_{\perp}$), $\chi_{\rm s}$ behaves in 
the same way for all layers and surprisingly leads
to a nonmonotonic $\alpha$-dependence for the center layer. 

The numerical data in Figs.~\ref{2lay} and \ref{3lay} show that
interlayer hopping is in competition with spatially modulated Rashba 
spin-orbit coupling, such that a larger
$t_{\perp}$ yields a higher effective $\alpha_{\rm eff}\sim t_{\perp}$
for the crossover from the behavior of a conventional superconductor to
that of a non-centrosymmetric superconductor. For $\alpha_{\rm eff}\gg
t_{\perp}$, layers are almost decoupled from each other (see \S4), 
and then the multilayer superconductor is regarded as a set of 
(non-)centrosymmetric superconductors. 
This crossover is best observed in the peak of
$\chi_{\rm s}$ at approximately $\alpha_{\rm eff}\sim t_{\perp}$ for the center
layer of the trilayer system (Fig.~\ref{3bunlay}). 
Thus, modifying $t_{\perp}$, e.g., by applying
uniaxial stress along the $c$-axis, can affect the magnetic response
for $c$-axis fields in case (B). No such effect is expected in case (A).  

\begin{figure}[htbp]
\begin{center}
\includegraphics[scale=0.35]{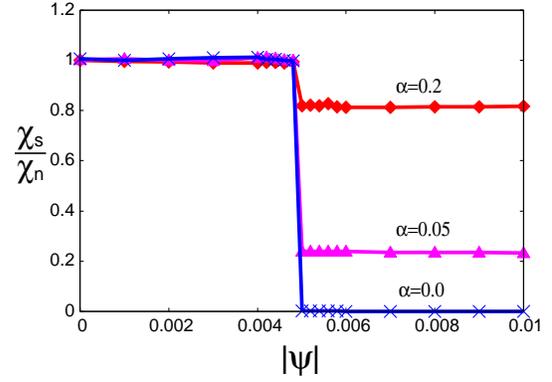}
\end{center}
\caption{(Color online) 
Zero-temperature spin susceptibility along $c$-axis for various $|\psi|$ values in 
bilayer system.  
We assume the amplitude of the spin triplet 
component to be $|d|=0.01 - |\psi|$. We show the results for interlayer 
coupling $t_{\perp}=0.1$ and spin-orbit coupling $\alpha=0$ (crosses), 
$\alpha=0.05$ (triangles), and $\alpha=0.2$ (diamonds).} 
\label{st2}
\end{figure}

Surprisingly, these results are almost independent of the subdominant 
order parameter. Figure~\ref{st2} shows the spin susceptibility of 
bilayers for various $|\psi|$ values while keeping the summation 
$|\psi| + |d| = 0.01$. We see a nearly constant spin susceptibility 
except for the jump at $|\psi| = |d| = 0.005$. 
This jump arises from a discontinuous change of the order parameter, 
since we assume case (A) for $|\psi| < |d|$ and case (B) 
for $|\psi| > |d|$. 
We will show that the spin susceptibility at $T=0$ 
is determined by the interlayer phase differences of order parameters 
(\S7). 
Since we assume the zero-phase difference for the dominant 
order parameter and the $\pi$-phase difference for the subdominant one, 
the spin susceptibility is determined by the dominant order parameter 
and is negligibly affected by the subdominant component. 
The details will be given in \S7.

\subsection{Spin susceptibility along $ab$-axis}

\begin{figure}[htbp]
\begin{center}
\includegraphics[scale=0.35]{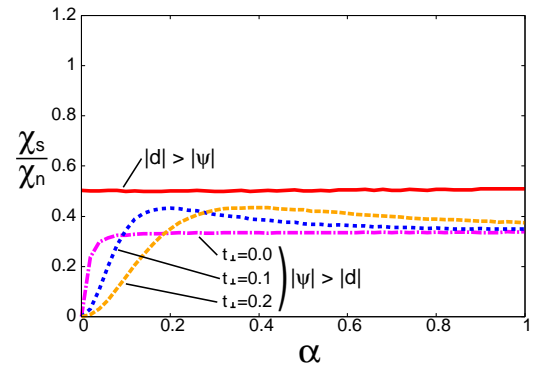}
\end{center}
\caption{(Color online) 
Spin susceptibility along $a$-axis for trilayer system.
Parameters are the same as those in Fig.~\ref{3lay}. 
}
\label{3xlay}
\end{figure}

Within our model, we find that the spin susceptibility along the
$ab$-axis is always half of that along the $c$-axis, independent of
the strength of $\alpha$ and $ t_{\perp} $ as well as of the number of 
layers. For numerical evidence, Fig.~\ref{3xlay} shows 
the spin susceptibility along the $ab$-axis in the trilayer system;
it is one-half of the results in Fig.~\ref{3lay}. 
We confirmed these behaviors for another number of layers. 

\section{Electronic Structure in Normal State}

 To elucidate the crossover from the behavior of a
centrosymmetric superconductor to that of a non-centrosymmetric 
superconductor shown in the numerical calculation 
(Figs.~\ref{2lay} and \ref{3lay}), 
we give an analytic expression of the single-particle wave function 
and superconducting gap in \S4 and \S5, respectively. 
 An intuitive understanding is given in \S6 by decomposing 
the spin susceptibility into the Pauli and Van-Vleck parts.

\subsection{Bilayer system}

\begin{figure}[htbp]
\begin{center}
\includegraphics[scale=0.35]{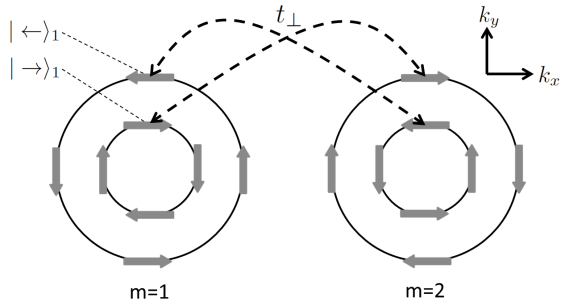}
\end{center}
\caption{Schematic figure of Fermi surfaces in bilayer system 
without interlayer coupling $t_\perp = 0$. 
In this case, layers are decoupled, and then the Fermi surfaces 
of layers 1 and 2 are independent of each other. 
The Fermi surfaces are split by Rashba spin-orbit coupling in 
both layers 1 and 2, but the spin orientations are opposite 
because Rashba spin-orbit coupling $\alpha_m$ has the opposite sign 
($\alpha_1 = - \alpha_2$). 
When interlayer coupling $t_\perp$ is tuned on, the electronic 
states are coupled, as shown by dashed arrows. 
}
\label{band}
\end{figure}

First, we study the single-particle wave function in the normal state 
of bilayer systems. 
The schematic figure for the role of Rashba spin-orbit coupling 
and interlayer coupling is shown in Fig.~\ref{band}. 
When the two layers are decoupled at $t_\perp = 0$, 
each layer has Fermi surfaces, as shown in Fig.~\ref{band}. 
The structures of Fermi surfaces are the same for layers 
1 and 2, but the spin orientations are opposite, because the 
Rashba spin-orbit coupling $\alpha_m$ has the opposite sign 
$\alpha_1 = -\alpha_2$. 
When the interlayer coupling $t_\perp$ is switched on, the Fermi 
surfaces are coupled, as shown by dashed arrows in Fig.~\ref{band}.

We show the wave function of quasi-particles with the momentum
$\k = (0,\ky)$ as an example. According to the Fermi surfaces shown 
in Fig.~\ref{band}, a simple expression is obtained 
by choosing the spin quantization axis along the $x$-direction.
In the following presentation, the wave functions 
$|\rightarrow\rangle_{m}$ and $|\leftarrow\rangle_{m}$ describe 
the electron states on the  layer $m$ with right- and 
left-pointing spins, respectively. 
In the normal state with $\Delta_{ss'm}(\mib{k})=0$, 
the Hamiltonian [eq.~(\ref{model})] is block-diagonalized 
in this representation. 
The block of the Hamiltonian is obtained as 
\begin{eqnarray}
\left(
\begin{array}{ccc}
\varepsilon(\k)+\alpha'(\k) & t_{\perp} \\
t_{\perp }& \varepsilon(\k)-\alpha'(\k) \\
\end{array}
\right)
\hspace*{1mm}
{\rm for }
\hspace*{1mm}
\left(
\begin{array}{ccc}
|\rightarrow\rangle_{1}\\
|\rightarrow\rangle_{2}\\
\end{array}
\right),
\end{eqnarray} 
and 
\begin{eqnarray}
\left(
\begin{array}{ccc}
\varepsilon(\k)-\alpha'(\k) & t_{\perp} \\
t_{\perp }& \varepsilon(\k)+\alpha'(\k) \\
\end{array}
\right)
\hspace*{1mm}
{\rm for }
\hspace*{1mm}
\left(
\begin{array}{ccc}
|\leftarrow\rangle_{1}\\
|\leftarrow\rangle_{2}\\
\end{array}
\right),
\end{eqnarray}
respectively. 
We denote the magnitude of spin-orbit coupling as
\begin{eqnarray}
\alpha'(\mib{k})\equiv\alpha|\mib{g}(\mib{k})|=\alpha\sqrt{\sin^{2}{k_{x}}+\sin^{2}{k_{y}}}.
\end{eqnarray}
Diagonalizing the $2 \times 2$ matrix, we obtain eigenvalues with 
twofold degeneracy: 
{\setlength\arraycolsep{1pt}
\begin{eqnarray}
\label{e21}
E^{(2)}_{1}(\mib{k})&=&\varepsilon(\mib{k})+\sqrt{\alpha'(\mib{k})^{2}+t_{\perp}^{2}}, \\
\label{e22}
E^{(2)}_{2}(\mib{k})&=&\varepsilon(\mib{k})-\sqrt{\alpha'(\mib{k})^{2}+t_{\perp}^{2}}.
\end{eqnarray}}
The wave function is obtained as 
\begin{eqnarray}
\left\{\ \begin{array}{l}
\sqrt{1-A(\k)^{2}}|\rightarrow\rangle_{1}+A(\k)|\rightarrow\rangle_{2} \\
A(\k)|\leftarrow\rangle_{1}+\sqrt{1-A(\k)^{2}}|\leftarrow\rangle_{2}
\end{array} \right.,
\label{WF21}
\end{eqnarray}
for $E=E^{(2)}_{1}(\k)$ and as
\begin{eqnarray}
\left\{\ \begin{array}{l}
-A(\k)|\rightarrow\rangle_{1}+\sqrt{1-A(\k)^{2}}|\rightarrow\rangle_{2} \\
\sqrt{1-A(\k)^{2}}|\leftarrow\rangle_{1}-A(\k)|\leftarrow\rangle_{2}
\end{array} \right.,
\label{WF22}
\end{eqnarray}
for $E=E^{(2)}_{2}(\k)$. 
We here defined   
\begin{eqnarray}
A(\mib{k}) \equiv \frac{t_{\perp}}{\sqrt{t_{\perp}^{2}
+\left(\alpha'(\k)+\sqrt{\alpha'(\k)^{2}+t_{\perp}^{2}}\right)^{2}}}.
\label{A-factor}
\end{eqnarray}
We obtain a similar single-particle wave function for the other 
momentum $\kx \ne 0$ by choosing a spin quantization axis 
parallel to $\mib{g}(\k)$.

The twofold degeneracy in the above two energy bands originates from 
the time-reversal and global inversion symmetry. 
Thus, the structure of the energy bands seems to be the same as that in 
conventional metals. 
However, Rashba spin-orbit coupling affects the single-particle 
wave function in a unique way. 
Note that $A(\k)$ in eq.~(\ref{A-factor}) is 
determined by the competition between interlayer coupling 
$t_\perp$ and spin-orbit coupling $\alpha$. 
When spin-orbit coupling is absent, i.e., $\alpha=0$, 
eq.~(\ref{A-factor}) is reduced to $A(\k)=1/\sqrt{2}$. 
Then, two energy bands are regarded as bonding and antibonding bands 
with spin degeneracy, namely, 
$(1/\sqrt{2})(|\rightarrow\rangle_{1} + |\rightarrow\rangle_{2})$ 
and 
$(1/\sqrt{2})(|\leftarrow\rangle_{1} +
|\leftarrow\rangle_{2})$ for $E=E^{(2)}_{1}(\k)$,  
and 
$(1/\sqrt{2})(|\rightarrow\rangle_{1} - |\rightarrow\rangle_{2})$ 
and 
$(1/\sqrt{2})(|\leftarrow\rangle_{1} - |\leftarrow\rangle_{2})$ 
for $E=E^{(2)}_{2}(\k)$, respectively. 
This is the conventional band structure of bilayer systems. 
On the other hand, we obtain $A(\k) = 0$ in the limit of large 
spin-orbit coupling $\alpha \gg t_\perp$, and then 
the twofold degeneracy is given by 
$|\rightarrow\rangle_{1}$ and $|\leftarrow\rangle_{2}$, 
and by $|\leftarrow\rangle_{1}$ and $|\rightarrow\rangle_{2}$. 
Thus, the quasi-particle is localized on each layer and degenerates 
with a quasi-particle of opposite spin on the other layer. 
In this case, the splitting of the two energy bands is induced by 
the spin splitting in each layer, although the twofold degeneracy 
is protected by the global inversion symmetry. 
With increasing spin-orbit coupling, the two energy bands change their 
character from the bonding and antibonding states to the spin split 
states on each layer. This crossover leads to the $\alpha$ dependence 
of spin susceptibility in Fig.~\ref{2lay}.

\subsection{Trilayer system}

\begin{figure}[htbp]
\begin{center}
\includegraphics[scale=0.35]{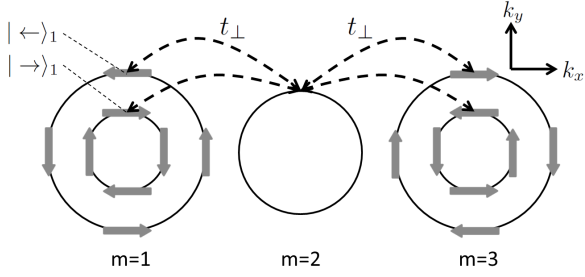}
\end{center}
\caption{Schematic figure of Fermi surfaces in trilayer system
without interlayer coupling $t_\perp = 0$. 
Fermi surfaces in layers 1 and 3 are split by Rashba 
spin-orbit coupling; however, they degenerate in layer 2 because 
of $\alpha_2 =0$.  
When interlayer coupling $t_\perp$ is turned on, the electronic 
states are coupled, as shown by dashed arrows. 
}
\label{band3}
\end{figure}

Next, we study the trilayer systems. 
The single-particle wave function is obtained 
in the same way as that in the bilayer system. 
Figure~\ref{band3} shows the schematic figure of Fermi surfaces in 
the decoupling limit $t_\perp =0$. 
For the quasi-particles with $\k = (0,\ky)$, the Hamiltonian without 
pairing field $\Delta_{ss'm}(\mib{k})=0$ is block-diagonalized as
{\setlength\arraycolsep{1pt}
\begin{eqnarray}
\left(
\begin{array}{ccc}
\varepsilon(\k)+\alpha'(\k) & t_{\perp} & 0 \\
t_{\perp} & \varepsilon(\k) & t_{\perp} \\
0 & t_{\perp } & \varepsilon(\k)-\alpha'(\k) \\
\end{array}
\right)
\hspace*{0.5mm}
{\rm for }
\hspace*{0.5mm}
\left(
\begin{array}{ccc}
|\rightarrow\rangle_{1}\\
|\rightarrow\rangle_{2}\\
|\rightarrow\rangle_{3}\\
\end{array}
\right),
\end{eqnarray}}
and as
{\setlength\arraycolsep{1pt}
\begin{eqnarray}
\left(
\begin{array}{ccc}
\varepsilon(\k)-\alpha'(\k) & t_{\perp} & 0 \\
t_{\perp} & \varepsilon(\k) & t_{\perp} \\
0 & t_{\perp } & \varepsilon(\k) +\alpha'(\k) \\
\end{array}
\right)
\hspace*{0.5mm}
{\rm for }
\hspace*{0.5mm}
\left(
\begin{array}{ccc}
|\leftarrow\rangle_{1}\\
|\leftarrow\rangle_{2}\\
|\leftarrow\rangle_{3}\\
\end{array}
\right).
\end{eqnarray}}
Diagonalizing the $3 \times 3$ matrix, we obtain three eigenvalues:  
{\setlength\arraycolsep{1pt}
\begin{eqnarray}
E^{(3)}_{1}(\mib{k})&=&\varepsilon(\mib{k})+\sqrt{\alpha'(\mib{k})^{2}+2t_{\perp}^{2}}, \\
E^{(3)}_{2}(\mib{k})&=&\varepsilon(\mib{k}), \\
E^{(3)}_{3}(\mib{k})&=&\varepsilon(\mib{k})-\sqrt{\alpha'(\mib{k})^{2}+2t_{\perp}^{2}}, 
\end{eqnarray}}
with twofold degeneracy. 
The wave function is described as 
\begin{eqnarray}
\left\{\ \begin{array}{l}
\sqrt{1-B^{2}-C^{2}}|\rightarrow\rangle_{1}+C|\rightarrow\rangle_{2}+B|\rightarrow\rangle_{3} \\
B|\leftarrow\rangle_{1}+C|\leftarrow\rangle_{2}+\sqrt{1-B^{2}-C^{2}}|\leftarrow\rangle_{3}
\end{array} \right.,
\label{WF31}
\end{eqnarray}
for $E=E^{(3)}_{1}(\k)$, 
\begin{eqnarray}
\left\{\ \begin{array}{l}
-C|\rightarrow\rangle_{1}+\sqrt{1-2C^{2}}|\rightarrow\rangle_{2}+C|\rightarrow\rangle_{3} \\
C|\leftarrow\rangle_{1}+\sqrt{1-2C^{2}}|\leftarrow\rangle_{2}-C|\leftarrow\rangle_{3}
\end{array} \right.,
\label{WF32}
\end{eqnarray}
for $E=E^{(3)}_{2}(\k)$, and
\begin{eqnarray}
\left\{\ \begin{array}{l}
B|\rightarrow\rangle_{1}-C|\rightarrow\rangle_{2}+\sqrt{1-B^{2}-C^{2}}|\rightarrow\rangle_{3} \\
\sqrt{1-B^{2}-C^{2}}|\leftarrow\rangle_{1}-C|\leftarrow\rangle_{2}+B|\leftarrow\rangle_{3}
\end{array} \right.,
\label{WF33}
\end{eqnarray}
for $E=E^{(3)}_{3}(\k)$. 
We omitted the index $\k$ in eqs.~(\ref{WF31})-(\ref{WF33}), and defined 
$B(\k)$ and $C(\k)$ as  
{\setlength\arraycolsep{1pt}
\begin{eqnarray}
\hspace*{-5mm}
B(\mib{k})&\equiv&\frac{t_{\perp}^{2}}{(\alpha'(\k)+\sqrt{\alpha'(\k)^{2}+2t_{\perp}^{2}})\sqrt{\alpha'(\k)^{2}+2t_{\perp}^{2}}}, 
\\ 
\hspace*{-5mm}
C(\mib{k})&\equiv&\frac{t_{\perp}}{\sqrt{\alpha'(\k)^{2}+2t_{\perp}^{2}}}, 
\end{eqnarray}}
respectively.

In the absence of spin-orbit coupling ($\alpha =0$), we obtain 
$B(\k) = 1/2$ and $C(\k) = 1/\sqrt{2}$; then 
the energy bands with $E=E^{(3)}_{1}(\k)$ and $E=E^{(3)}_{3}(\k)$
have an even parity with respect to the center layer, 
while the band $E=E^{(3)}_{2}(\k)$ has the odd parity. 
When the spin-orbit coupling is turned on, the parity is mixed. 
In the limit of large spin-orbit coupling $\alpha \gg t_\perp$,
the quasi-particle is localized on each layer, as shown in 
Fig.~\ref{band3}.  
The doubly degenerate wave functions are 
$|\rightarrow\rangle_{1}$ and $|\leftarrow\rangle_{3}$ 
for $E=E^{(3)}_{1}(\k)$, 
$|\rightarrow\rangle_{2}$ and $|\leftarrow\rangle_{2}$ 
for $E=E^{(3)}_{2}(\k)$, and 
$|\rightarrow\rangle_{3}$ and $|\leftarrow\rangle_{1}$ 
for $E=E^{(3)}_{3}(\k)$.

\section{Order Parameter in Band Basis}

We turn to a superconducting order parameter in the band basis. 
For this purpose, the Hamiltonian is transformed by the unitary matrix 
$T(\k)$, which diagonalizes the Hamiltonian without a pairing field 
($\Delta_{ss'm}(\mib{k})=0$),  
{\setlength\arraycolsep{1pt}
\begin{eqnarray}
H = \frac{1}{2} \sum_{\k} \hat{C}_{\k}^{\dag} \hat{H}(\k) \hat{C}_{\k}
= \frac{1}{2} \sum_{\k} \hat{Z}_{\k}^{\dag} \hat{H}'(\k) \hat{Z}_{\k}. 
\end{eqnarray}}
The basis of $\hat{Z}_{\k}^{\dag}$ is chosen to be the eigenstates in 
eqs.~(\ref{WF21}) and (\ref{WF22}) for bilayers. 
Then, we obtain the following matrix representation of the Hamiltonian;
{\setlength\arraycolsep{1pt}
\begin{eqnarray}
&& \hspace*{-5mm}
\hat{H}'(\k) = 
\hat{T}^{\dag}(\mib{k})\hat{H}(\mib{k})\hat{T}(\mib{k}) \nonumber \\
&=&\left(
\begin{array}{cccccccc}
E^{(2)}_{1} & 0 & 0 & 0 & \Delta_{1} & 0 & \Delta_{13} & 0 \\
0 & E^{(2)}_{1} & 0 & 0 & 0 & \Delta_{2} & 0 & \Delta_{24} \\
0 & 0 & E^{(2)}_{2} & 0 & \Delta_{31} & 0 & \Delta_{3} & 0 \\
0 & 0 & 0 & E^{(2)}_{2} & 0 & \Delta_{42} & 0 & \Delta_{4} \\
\Delta_{1}^{\ast} & 0 & \Delta_{31}^{\ast} & 0 & -E^{(2)}_{1} & 0 & 0 & 0 \\
0 & \Delta_{2}^{\ast} & 0 & \Delta_{42}^{\ast} & 0 & -E^{(2)}_{1} & 0 & 0 \\
\Delta_{13}^{\ast} & 0 & \Delta_{3}^{\ast} & 0 & 0 & 0 & -E^{(2)}_{2} & 0 \\
0 & \Delta_{24}^{\ast} & 0 & \Delta_{4}^{\ast} & 0 & 0 & 0 & -E^{(2)}_{2}
\end{array}\right), \nonumber \\
\label{Delta_band}
\end{eqnarray}}
where the index $\k$ is omitted for simplicity. 

Since we consider a small superconducting gap $|\Delta| \ll |\alpha|$, 
as realized in most (locally) non-centrosymmetric 
superconductors, the role of interband Cooper pairing is ignored. 
Then, the superconducting order parameter of each band is described by 
$\Delta_{i}(\k)$ because the $2 \times 2$ matrix for 
intraband Cooper pairing is diagonalized in this representation. 
Thus, the superconductivity in bilayers is regarded as 
an equal pseudo-spin pairing state on the basis of 
eqs.~(\ref{WF21}) and (\ref{WF22}). 

The superconducting order parameter in each band is obtained as 
follows. There are two cases with respect to the interlayer phase 
difference, as discussed in \S2. 
In case (A), the spin triplet order parameter (d-vector) 
has the same sign in both layers 1 and 2, while the spin singlet 
order parameter has opposite signs in the two layers, namely, 
$(\psi_1, \psi_2) = (\psi, -\psi)$ and $(d_1, d_2) = (d, d)$. 
In case (B), the spin singlet order parameter has the same 
sign in layers 1 and 2, but the d-vector has opposite signs in the two layers, 
namely, $(\psi_1, \psi_2) = (\psi, \psi)$ and $(d_1, d_2) = (d, -d)$.

Case (A) leads to the order parameter in the band basis
for $\k =(0,\ky)$: 
{\setlength\arraycolsep{1pt}
\begin{eqnarray}
\Delta_{1}(\mib{k})&=&-k_{-}\left[\frac{\alpha}{\sqrt{\alpha^{2}|\mib{g}(\mib{k})|^{2}+t_{\perp}^{2}}}\psi+d\right], 
\label{p1}
\\
\Delta_{2}(\mib{k})&=&\Delta_{1}(\mib{k}), 
\label{p2}
\\
\Delta_{3}(\mib{k})&=&-k_{-}\left[\frac{\alpha}{\sqrt{\alpha^{2}|\mib{g}(\mib{k})|^{2}+t_{\perp}^{2}}}\psi-d\right], 
\label{p3}
\\
\Delta_{4}(\mib{k})&=&\Delta_{3}(\mib{k}),
\label{p4}
\end{eqnarray}}
where $k_{\pm}\equiv\sin{k_{y}}\pm{\rm{i}}\sin{k_{x}}$. 
On the other hand, we obtain the following results in case (B): 
{\setlength\arraycolsep{1pt}
\begin{eqnarray}
\Delta_{1}(\mib{k})&=&\frac{k_{-}}{|\mib{g}(\mib{k})|}\left[\psi+\frac{\alpha|\mib{g}(\mib{k})|^{2}}{\sqrt{\alpha^{2}|\mib{g}(\mib{k})|^{2}+t_{\perp}^{2}}}d\right], 
\label{d1} 
\\
\Delta_{2}(\mib{k})&=&-\Delta_{1}(\mib{k}), \label{d2}\\
\Delta_{3}(\mib{k})&=&-\frac{k_{-}}{|\mib{g}(\mib{k})|}\left[\psi-\frac{\alpha|\mib{g}(\mib{k})|^{2}}{\sqrt{\alpha^{2}|\mib{g}(\mib{k})|^{2}+t_{\perp}^{2}}}d\right], 
\label{d3}
\\
\Delta_{4}(\mib{k})&=&-\Delta_{3}(\mib{k}). 
\label{d4}
\end{eqnarray}}
Thus, the relative sign of $\Delta_1 (\k)$ and $\Delta_2 (\k)$ 
and that of $\Delta_3 (\k)$ and $\Delta_4 (\k)$ are different 
in cases (A) and (B) and thus also yield different spin susceptibilities, as shown
in Fig.~\ref{2lay} and discussed in \S7. 
We do not show the superconducting order parameter in the trilayer 
systems, but it is straightforward to extend our analytical calculation to trilayers.

\section{Spin Susceptibility}

\subsection{Pauli and Van-Vleck contributions}

We provide here the theoretical basis 
of our numerical results of 
spin susceptibility in \S3. For this purpose, we decompose 
the spin susceptibility in the normal state 
into the Pauli and Van-Vleck parts. 
With the use of the linear response theory, the transverse component of 
uniform spin susceptibility is obtained in the normal state as 
{\setlength\arraycolsep{1pt}
\begin{eqnarray}
\chi^{+-}_{\rm n}&=&\lim_{\mib{q} \to
 0}\sum_{\eta,\nu}\sum_{\mib{k}}\langle\eta |
 S^{+} |\nu\rangle\langle\nu |
 S^{-}|\eta\rangle \nonumber 
\\
 &&\times\frac{f(E_{\eta}(\mib{k}))-f(E_{\nu}(\mib{k}+\mib{q}))}{E_{\nu}(\mib{k}+\mib{q})-E_{\eta}(\mib{k})}, 
\label{tz}
\end{eqnarray}}
which is decomposed into  
\begin{eqnarray}
\chi^{+-}_{\rm n}=\chi^{\rm{P}}+\chi^{\rm{V}}.
\end{eqnarray}
The Pauli spin susceptibility $\chi^{\rm{P}}$ arises from 
the intraband contributions, while the interband polarization gives rise 
to the Van-Vleck spin susceptibility $\chi^{\rm{V}}$. They can be expressed as
{\setlength\arraycolsep{1pt}
\begin{eqnarray}
\chi^{\rm{P}}&=&\sum_{E_{\eta}=E_{\nu}}\sum_{\mib{k}}\langle\eta |
 S^{+} |\nu\rangle\langle\nu |
 S^{-}|\eta\rangle\delta(E_{\eta}(\mib{k})),
\label{pa}
\\
\chi^{\rm{V}}&=&\sum_{E_{\eta}\neq E_{\nu}}\sum_{\mib{k}}\langle\eta |
 S^{+} |\nu\rangle\langle\nu |
 S^{-}|\eta\rangle 
\frac{f(E_{\eta}(\mib{k}))-f(E_{\nu}(\mib{k}))}{E_{\nu}(\mib{k})-E_{\eta}(\mib{k})}, \nonumber \\
\label{va}
\end{eqnarray}}
respectively. 
In multilayer systems, the Fermi surfaces are split by both 
spin-orbit coupling and interlayer coupling, whereby the former
gives rise to the Van-Vleck susceptibility.

For the discussion of the spin susceptibility on the basis of 
the single-particle wave functions given in \S4, it should be noted that the 
spin quantization axis is directed along the $ab$-plane
in the entire Brillouin zone. 
Since the spin quantization axis is perpendicular to the $c$-axis, 
the spin susceptibility along the $c$-axis is given by the transverse 
spin component that can be calculated using eq.~(\ref{tz}). 

The Van-Vleck spin susceptibility is negligibly affected by the 
superconducting gap when the band splitting due to the spin-orbit 
coupling and interlayer coupling is much larger than the 
superconducting gap. 
Thus, the Van-Vleck contribution provides a lower limit of spin 
susceptibility in the superconducting state. 
On the other hand, the Pauli spin susceptibility may be suppressed by 
the superconductivity and depends strongly on the symmetry of the order parameter. 
We will show that the Pauli spin susceptibility of multilayer 
superconductors is determined by the phase difference of the order parameter, 
which has been discussed for bilayers in \S5, irrespective of the 
ratio of the spin singlet and the triplet order parameters. 
The Pauli spin susceptibility is completely suppressed in case (B), 
whereas it remains unaffected in case (A), 
as will be shown in \S7. 
 
We assumed case (A) [case (B)] in the dominant spin triplet 
(singlet) pairing state for the numerical calculation in \S3. 
Thus, the spin susceptibility in the 
dominantly spin singlet superconducting state studied in \S3 coincides 
with the Van-Vleck part in the normal state, and therefore 
$\chi_{\rm s}/\chi_{\rm n} = 
\chi^{\rm{V}}/(\chi^{\rm{P}}+\chi^{\rm{V}})$, but 
$\chi_{\rm s}/\chi_{\rm n} =1$ in the dominantly spin triplet 
superconducting state. 
We discuss the Pauli and Van-Vleck spin susceptibilities 
of bi- and tri-layers below. 

\subsection{Bilayer system}

With the use of the single-particle wave functions eqs.~(\ref{WF21}) 
and (\ref{WF22}) in the bilayer system, we obtain 
the normal state Pauli and Van-Vleck spin susceptibilities at $T=0$ as  
\begin{eqnarray}
\label{chi-P-bi}
&& \hspace*{-10mm} 
\chi^{\rm P} 
=\sum_{\mib{k}}4A(\k)^{2}\bigr(1-A(\k)^{2}\bigr) \,
[\delta(E^{(2)}_{1}(\k))+\delta(E^{(2)}_{2}(\k))] \nonumber \\
\\ &&  \hspace*{-5mm}
\sim \rho(0) \langle \, 4A(\k)^{2}\bigr(1-A(\k)^{2}\bigr) \, \rangle_{\rm FS}, 
\end{eqnarray}
and 
\begin{eqnarray}
\label{chi-V-bi}
&& \hspace*{-14mm}
\chi^{\rm V}
=\sum_{\mib{k}} 2\bigr(1-2A(\k)^{2}\bigr)^{2} \,\,
\frac{\Theta(E^{(2)}_{1}(\k))-\Theta(E^{(2)}_{2}(\k))}
{E^{(2)}_{2}(\k)-E^{(2)}_{1}(\k)} 
\\ && \hspace*{-9mm}
\sim \rho(0) \langle \, \bigr(1-2A(\k)^{2}\bigr)^{2} \, \rangle_{\rm FS},
\end{eqnarray}
respectively. 
Note that the Fermi-Dirac distribution function $f(E)$ is reduced to 
the step function $\Theta(E)$ at $T=0$. 
The average on the Fermi surface is denoted as 
$\langle\cdots\rangle_{\rm FS}$ and $\rho(0)$ is 
the density of state at the Fermi level. 

In the absence of spin-orbit coupling, the Van-Vleck part vanishes 
and the Pauli part is obtained as $\chi^{\rm P} = \chi^{+-}_{\rm n}$ 
because of $A(\k) = 1/\sqrt{2}$. 
On the other hand, the Pauli part vanishes for $t_\perp =0$ 
such that $\chi^{\rm V} = \chi^{+-}_{\rm n}$ because of $A(\k) = 0$.

\begin{figure}
\begin{center}
\includegraphics[scale=0.35]{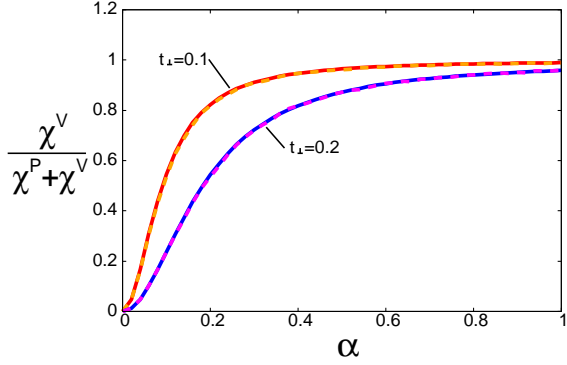}
\end{center}
\caption{(Color online) 
Van-Vleck part of spin susceptibility along $c$-axis in 
normal state of bilayer system, 
$\chi^{\rm V}/(\chi^{\rm P}+\chi^{\rm V})$ (solid lines). 
The dashed lines show the spin susceptibility in the superconducting 
state, $\chi_{\rm s}/\chi_{\rm n}$, for $\psi = 0.01$ and $d = 0$, 
which is shown in Fig.~\ref{2lay}.  
We see that these quantities coincide with each other. 
The parameter is chosen as $t_{\perp} = 0.1$ or $0.2$. 
}
\label{pv2n}
\end{figure}

We show the numerical results of 
$\chi^{\rm{V}}/(\chi^{\rm{P}}+\chi^{\rm{V}})$ obtained using 
eqs.~(\ref{chi-P-bi}) and (\ref{chi-V-bi}) 
for various spin-orbit couplings $\alpha$ in Fig.~\ref{pv2n}. 
We see that the Van-Vleck part of the spin susceptibility in the normal state, 
$\chi^{\rm V}/(\chi^{\rm P}+\chi^{\rm V})$, 
increases with $\alpha$ and coincides with the spin susceptibility, 
$\chi_{\rm s}/\chi_{\rm n}$, in the superconducting state with a dominant 
spin singlet order parameter. 
Thus, we obtain an estimate of the spin susceptibility in the superconducting 
state through the Van-Vleck spin susceptibility in the normal state.

\subsection{Trilayer system}

With the use of the single-particle wave functions 
eqs.~(\ref{WF31})-(\ref{WF33}) in the trilayers,
the normal state Pauli spin susceptibility at $T=0$ 
is obtained as  
{\setlength\arraycolsep{1pt}
\begin{eqnarray}
&& \hspace*{-2mm}
\chi^{\rm P}= 
\sum_{\mib{k}} \biggr\{ 
(2B\sqrt{1-B^{2}-C^{2}}+C^{2})^{2}
[\delta(E^{(3)}_{1})+\delta(E^{(3)}_{3})]
\nonumber \\ && \hspace*{8mm}
+\bigr(1-4C^{2}\bigr)^{2} \delta(E^{(3)}_{2}) \biggr\} 
\label{chi-P-tri}
\\ && \hspace*{2mm}
\sim \frac{2\rho(0)}{3} 
\langle 
(2B\sqrt{1-B^{2}-C^{2}}+C^{2})^{2} 
\rangle_{\rm FS}
\nonumber 
\\ && \hspace*{6mm}
+ \frac{\rho(0)}{3} 
\langle 
\bigr(1-4C^{2}\bigr)^{2}
\rangle_{\rm FS}, 
\end{eqnarray}}
while the Van-Vleck spin susceptibility is obtained as  
{\setlength\arraycolsep{1pt}
\begin{eqnarray}
\chi^{\rm V}&=&\sum_{\mib{k}}\Bigr\{2C^{2} 
\bigr(\sqrt{1-B^{2}-C^{2}}
+\sqrt{1-2C^{2}}-B \bigr)^{2}
\nonumber 
\\ && \hspace*{3mm}
\times\left[\frac{\Theta(E^{(3)}_{1})-\Theta(E^{(3)}_{2})}
{E^{(3)}_{2}-E^{(3)}_{1}}
+\frac{\Theta(E^{(3)}_{2})-\Theta(E^{(3)}_{3})}
{E^{(3)}_{3}-E^{(3)}_{2}}\right] 
\nonumber 
\\ && \hspace*{3mm}
+2 \bigr(1-2C^{2} \bigr)^{2}
\frac{\Theta(E^{(3)}_{1})-\Theta(E^{(3)}_{3})}
{E^{(3)}_{3}-E^{(3)}_{1}}\Bigr\}
\label{chi-V-tri}
\\ && \hspace*{-2mm}
\sim  
\frac{2\rho(0)}{3}
\langle 
2C^{2} 
\bigr(\sqrt{1-B^{2}-C^{2}}
+\sqrt{1-2C^{2}}-B \bigr)^{2} 
\rangle_{\rm FS}
\nonumber 
\\ && \hspace*{3mm}
+ \frac{\rho(0)}{3}
\langle 
2 \bigr(1-2C^{2} \bigr)^{2}
\rangle_{\rm FS}, 
\end{eqnarray}}
where we omitted the index $\k$ in $E^{(3)}_i(\k)$, $B(\k)$, and $C(\k)$ 
for simplicity. 

\begin{figure}
\begin{center}
\includegraphics[scale=0.35]{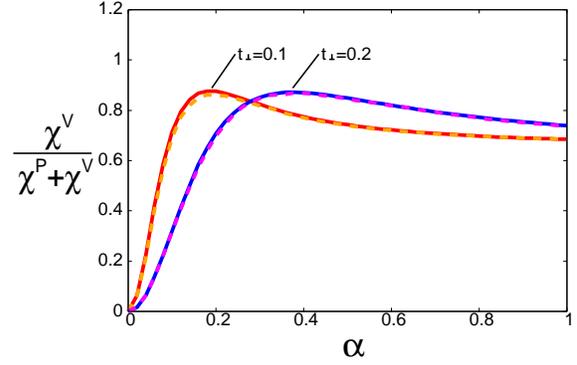}
\end{center}
\caption{(Color online) 
Van-Vleck part of spin susceptibility along $c$-axis in 
normal state of trilayer system, 
$\chi^{\rm V}/(\chi^{\rm P}+\chi^{\rm V})$ (solid lines). 
The dashed lines show the spin susceptibility in the superconducting 
state, $\chi_{\rm s}/\chi_{\rm n}$, for $\psi = 0.01$ and $d = 0$, 
which is shown in Fig.~\ref{3lay}.  
We see that these quantities coincide with each other, as in the 
bilayer systems. 
The parameter is chosen as $t_{\perp} = 0.1$ or $0.2$. 
}
\label{pv3n}
\end{figure}

The Van-Vleck spin susceptibility vanishes in the absence of the spin-orbit 
coupling $\alpha=0$, and 
$\chi^{\rm V} \sim 2\rho(0)/3 = 2\chi_{\rm n}/3$ 
in the limit of large spin-orbit coupling $\alpha \gg t_\perp$, 
as expected from the discussions in \S3 and \S6.1. 
Figure~\ref{pv3n} shows that the Van-Vleck part of the spin susceptibility 
in the normal state $\chi^{\rm{V}}/(\chi^{\rm{P}}+\chi^{\rm{V}})$ given 
by eqs.~(\ref{chi-P-tri}) and (\ref{chi-V-tri}) 
coincides with the spin susceptibility in the 
superconducting state $\chi_{\rm s}/\chi_{\rm n}$, which is shown
in Fig.~\ref{3lay}. 
Figure~\ref{pv3n} shows that the Pauli spin susceptibility is 
completely suppressed in the dominant spin singlet pairing state; 
therefore, the spin susceptibility of trilayers at $T=0$ 
is given by the Van-Vleck part, as in the bilayer system.

\section{Relation to Symmetry of Superconductivity}

In this section, we examine the relation between the order parameter 
and spin susceptibility in our multilayer superconductors. 
As we discussed in \S6, the Van-Vleck part of the spin susceptibility 
is not affected by superconductivity. On the other hand, 
the Pauli spin susceptibility has a clear dependence on the superconducting 
order parameter. Indeed, as we will show later,
the Pauli spin susceptibility is determined by the 
phase difference of the order parameter between layers, 
and is essentially independent of the ratio of parity mixing, $|\psi|/|d|$.

For the bilayer system, we consider cases (A) and (B), discussed in \S5, 
to optimize the interlayer Josephson coupling energy.  

To understand the above-described aspect of the order parameter structure, it is 
important to view the situation in the band basis, as in eqs.~(\ref{p1})-(\ref{d4}). 
For $\k = (0,\ky)$, in the band basis, the ''d-vector'' of the pseudo-spin  
points along the $y$-axis in case (A) because 
$\Delta_{1}(\k) = \Delta_{2}(\k)$ and $\Delta_{3}(\k) = \Delta_{4}(\k)$. 
On the other hand, in case (B) the ''d-vector'' is oriented along the $x$-axis, 
whereby we find
$\Delta_{1}(\k) = -\Delta_{2}(\k)$ and $\Delta_{3}(\k) = -\Delta_{4}(\k)$.
Since the spin quantization axis is along the $a$-axis of tetragonal 
crystals for $\k = (0,\ky)$, the crystal $c$-axis corresponds to the 
$y$-axis with the $z$-axis being the spin quantization axis. 
Thus, the ''d-vector'' of a pseudo-spin is parallel to the $c$-axis 
in case (A) and perpendicular to the $c$-axis in case (B). 
This is true not only for $\k = (0,\ky)$ but also for all momenta in the
Brillouin zone. 
One may mistakenly conclude that the Pauli spin susceptibility 
for fields along the $c$-axis is suppressed (unchanged) in case (A) 
[case (B)] similarly to a spin triplet 
superconductor with a d-vector parallel (perpendicular) to the field. 
However, the opposite relation is actually obtained for the 
''d-vector'' in the band basis, because the spin of each band is 
inversely polarized for $\k$ and $-\k$. 
Therefore, taking into account the opposite sign of the spin quantization axis, 
it is shown that the Pauli spin susceptibility along the $c$-axis 
remains unchanged by the superconductivity in case (A), while 
it is completely suppressed in case (B). 

We conclude here that the spin susceptibility is 
determined by the interlayer phase difference of 
the order parameters and independent 
of their amplitude. This surprising result has already been anticipated from  
Fig.~\ref{st2}. The spin susceptibility is nearly constant for $\psi$, 
although it shows a jump at $|\psi| = |d|$, following the discontinuity 
of the phase difference. 
Note that we assumed case (A) for $|\psi| < |d|$ and 
case (B) for $|\psi| > |d|$. 

\begin{figure}[htbp]
  \begin{center}
     \includegraphics[width=6cm]{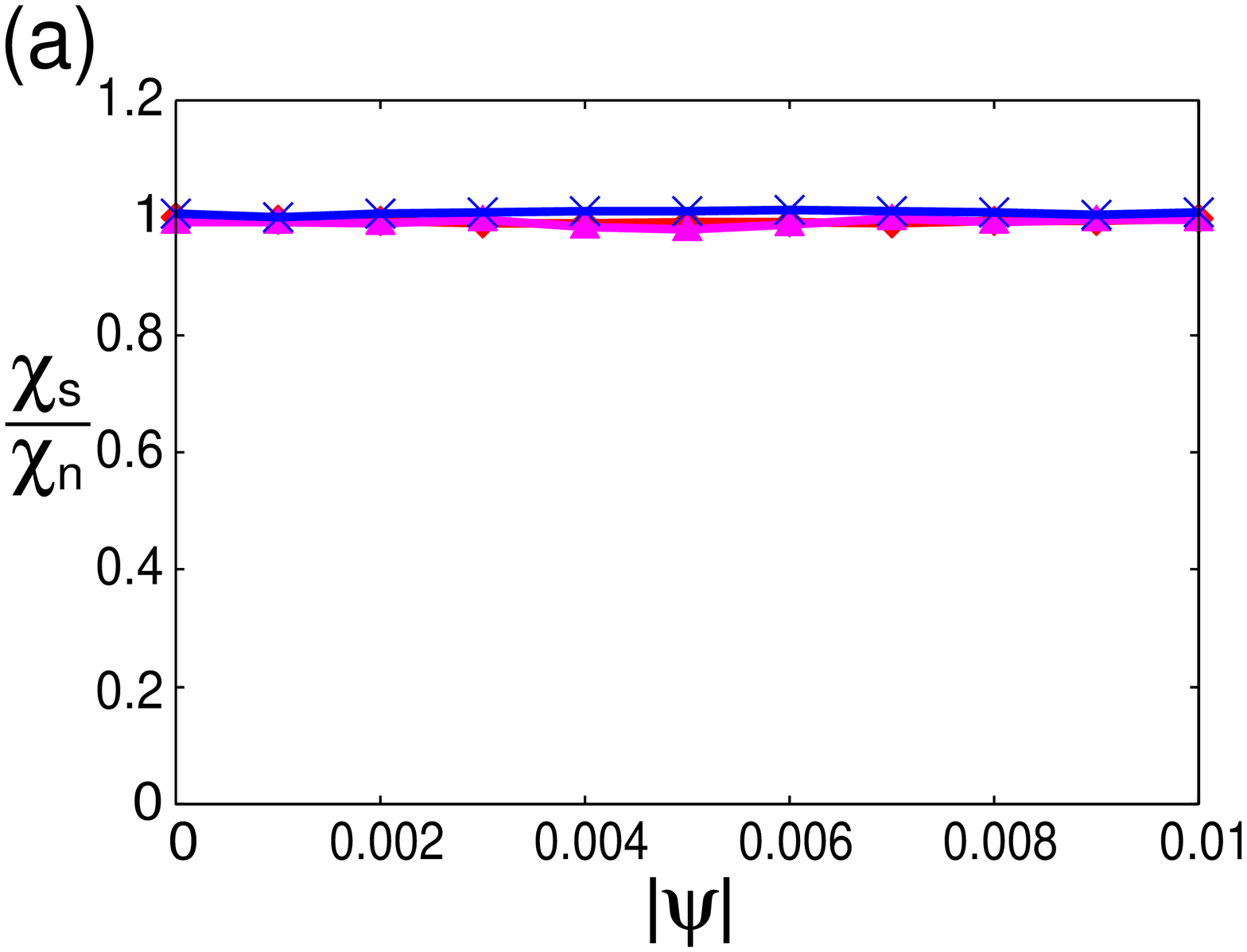}
     \includegraphics[width=6cm]{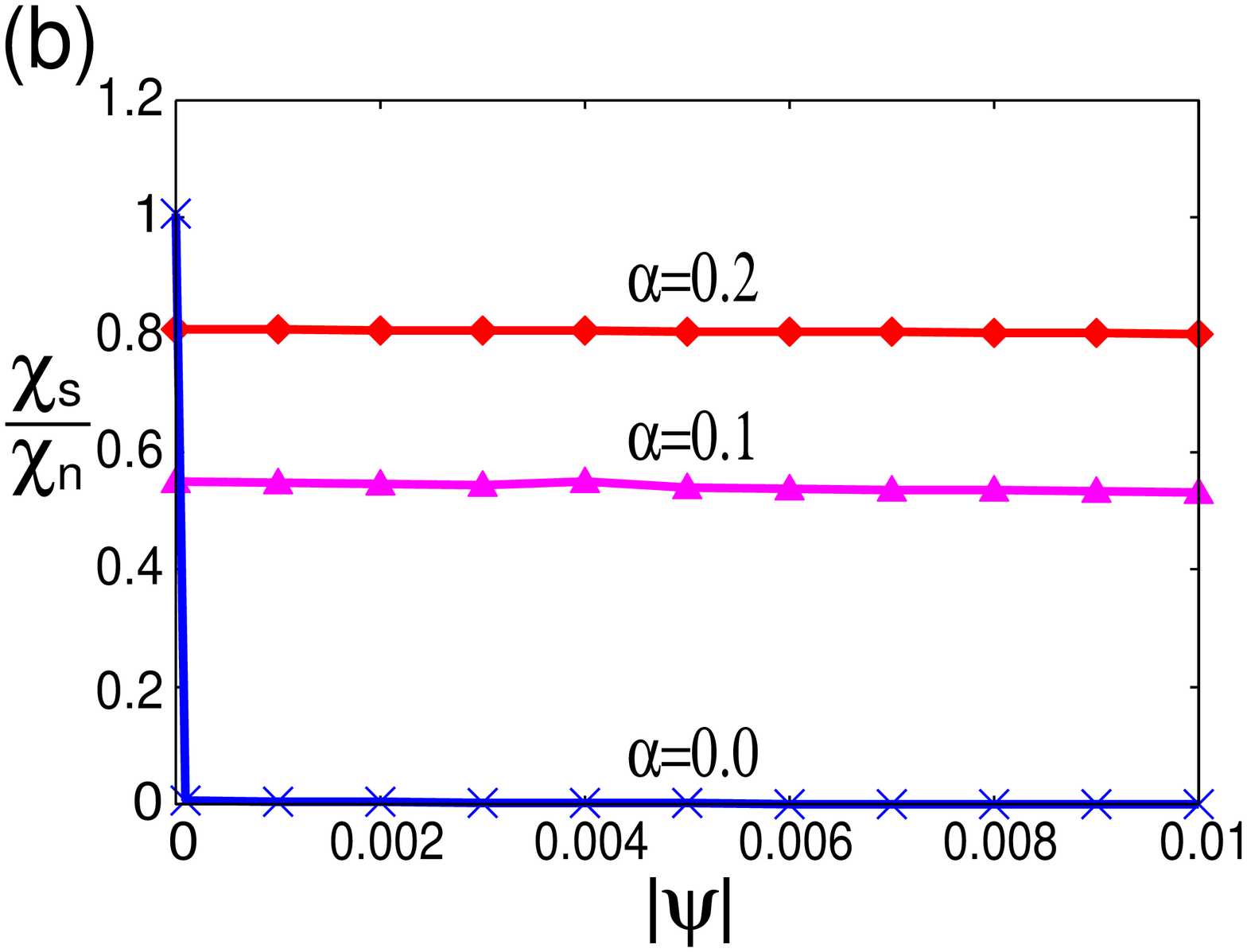}
  \end{center}
  \caption{(Color online) 
$C$-axis spin susceptibility in bilayer system 
for various $|\psi|$ values while keeping summation $|\psi| + |d| = 0.01$. 
The ratio $|\psi|/|d|$ changes from $0$ to $\infty$. 
We fix the phase difference of the order parameters for the two layers, 
irrespective of the ratio $|\psi|/|d|$, 
although interlayer Josephson coupling favors the zero 
phase difference of the dominant order parameter, as in \S3. 
(a) Case (A) in which spin singlet 
order parameter has a $\pi$-phase difference between layers.
(b) Case (B) in which spin singlet 
order parameter has a zero-phase difference.
We show the results for the interlayer coupling $t_{\perp}=0.1$ and the 
spin-orbit coupling $\alpha=0$ (crosses), $\alpha=0.1$ (triangles), 
and $\alpha=0.2$ (diamonds). 
}
\label{gpd}
\end{figure}

The above finding is confirmed by calculating the spin susceptibility 
while assuming case (A) or (B) regardless of the amplitude 
of singlet and triplet order parameters. 
Figure~\ref{gpd}(a) shows the result in case (A), while Fig.~\ref{gpd}(b) shows 
that in case (B). 
It is shown that the spin susceptibility at $T=0$ is nearly 
independent of the ratio of parity mixing from $|\psi|/|d| =0$ 
to $|\psi|/|d| = \infty$. 
Thus, the spin susceptibility remains unchanged when the dominant 
order parameter is a spin singlet component with a $\pi$-phase 
difference. On the other hand, the Pauli spin susceptibility is completely 
suppressed by the spin triplet order parameter with a $\pi$-phase 
difference between layers. 
In other words, the spin singlet order parameter with a $\pi$-phase 
difference plays the role of a spin triplet order parameter with a zero-phase difference, 
and {\it vice versa}. 

We obtain the same conclusion for the trilayer systems. 
The spin susceptibility at $T=0$ is determined by the phase difference 
and is independent of the amplitude of the order parameters. 
We confirmed that the spin susceptibility of trilayer superconductors 
is nearly constant for $|\psi|/|d|$, as in Fig.~\ref{gpd}.

\section{More than Three Layers}

Finally, we show the numerical results of the spin susceptibility 
for systems with 4 , 5 , 6, and 7 layers.  
Superconductivity has been observed on the artificial superlattice of 
CeCoIn$_5$ with 3, 4, 5, 7, and 9 layers~\cite{private.superlattices}. 
We consider the dominantly spin singlet pairing state 
with $\psi_{m}=0.01$ and ignore the spin triplet component. 
The subdominant component negligibly changes the spin susceptibility, 
as in the bi- and tri-layer systems. 

\begin{figure}[htbp]
\begin{center}
\includegraphics[scale=0.35]{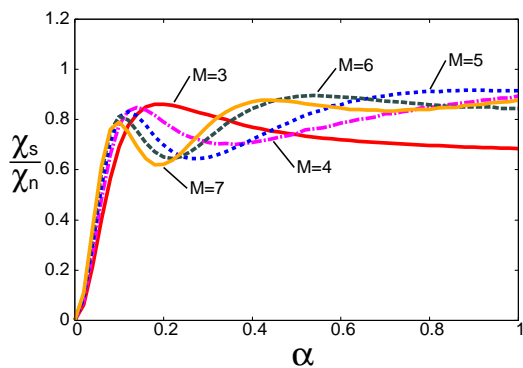}
\end{center}
\caption{(Color online) 
Spin susceptibilities along $c$-axis for $M=3,4,5,6,$ and $7$. 
We assume a weak screening effect of Rashba spin-orbit coupling, 
as explained in the text. 
We assume the uniform spin singlet superconducting order parameter 
$\psi_{m}=0.01$ and the interlayer coupling $t_{\perp}=0.1$. 
}
\label{c37}
\end{figure}

In this situation, it is necessary to extend the layer dependence of
the Rashba spin-orbit coupling, which is assumed to be weaker but still existent
in inner layers. For illustration, we choose two distinct layer dependences
of spin-orbit coupling, as we have no {\it ab initio} basis for our model.  
First, we consider a slow reduction (screening) of spin-orbit coupling, assuming
a decay proportional to the inverse square of layer distance. Such a model yields 
$(\alpha_{1}, \alpha_{2}, \alpha_{3}, \alpha_{4}) = \alpha (1,9/49,-9/49,-1)$ 
for the 4-layer system, and 
$(\alpha_{1}, \alpha_{2}, \alpha_{3}, \alpha_{4}, \alpha_{5}) 
= \alpha (1,9/41,0,-9/41,-1)$ for the 5-layer system. 
On the basis of this kind of model, we obtain the susceptibilities for different multilayer
systems, as shown in Fig.~\ref{c37}. We observe a nonmonotonic $\alpha$-dependence 
of the spin susceptibility because, in our model, spin-orbit coupling has 
several crossover scales (marked by several maxima) at $\alpha_1 \sim t_\perp$, 
$\alpha_2 \sim t_\perp$, and so on. 
For large $\alpha \gg t_\perp$, $\chi_{\rm s}$ recovers the normal 
state value $\chi_{\rm n}$ for even numbers of layers, 
such as 4 and 6, whereas  $\chi_{\rm s}/\chi_{\rm n} \rightarrow (M-1)/M$ 
for odd numbers of layers, because the center layer has no 
Rashba spin-orbit coupling. 

\begin{figure}[htbp]
\begin{center}
\includegraphics[scale=0.35]{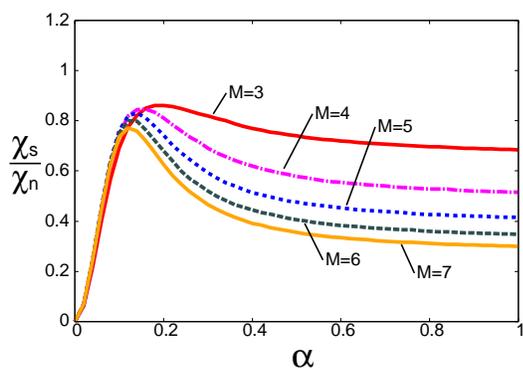}
\end{center}
\caption{(Color online) 
Spin susceptibility along $c$-axis in more than three layer systems 
with strong screening 
effect of Rashba spin-orbit coupling. Details are explained in the text. 
The other parameters are the same as those in Fig.~\ref{c37}. 
}
\label{c37a0}
\end{figure}

Next, we consider an extreme situation in which Rashba spin-orbit coupling 
only exists for the outermost layers, i.e., $\alpha_{m}=0$ 
except the outer layers $\alpha_1=\alpha$ and $\alpha_{M}=-\alpha$. 
In this case, the spin susceptibility shows a single peak at approximately 
$\alpha \sim t_\perp$ (Fig.~\ref{c37a0}), as in the trilayer systems. 
For large spin-orbit coupling $\alpha \gg t_\perp$, the spin 
susceptibility approaches $\chi_{\rm s}/\chi_{\rm n} = 2/M$.

Figures~\ref{c37} and \ref{c37a0} show nontrivial behaviors of the
spin susceptibility for multilayer systems, which is affected qualitatively 
by the layer dependence of Rashba spin-orbit coupling. 
Recently, the magnetic properties of superconducting 
multilayer cuprates have been investigated by NMR 
measurements.~\cite{Mukuda,Shimizu-Mukuda} 
In view of our discussion above, it would be interesting to perform
similar NMR measurements to
study the local spin polarization and to identify the effect of
symmetry-induced spin-orbit coupling in multilayer superconductors.

\section{Summary and Discussion}

Motivated by the recent investigation of artificially fabricated
superstructures of CeCoIn$_5$/YbCoIn$_5$, we have 
theoretically analyzed the basic properties of 
{\it locally} non-centrosymmetric superconductors in which 
spatially modulated Rashba spin-orbit coupling plays an 
important role. 
The single-particle wave function, superconducting gap, 
and spin susceptibility in the superconducting state have been determined.  

Although these layered systems possess a center of inversion symmetry,
they display local non-centrosymmetricity that affects the 
spin polarizability of the superconducting phase in an interesting
way. We clearly observe distinct regimes for the response of
such a layered superconductor with strong or weak interlayer
coupling. Although the former case exhibits a rather conventional response
in the spin polarization to an external field, the latter reflects properties
close to those expected for non-centrosymmetric superconductors. 
An intuitive understanding of the crossover between the two regimes
is obtained if we decompose the spin susceptibility into the conventional Pauli 
and an additional Van-Vleck contributions. The latter results from a spin-orbit
coupling-induced interband contribution, which is only weakly modified 
by superconductivity, while the Pauli part, as an intraband contribution, 
depends on details of the superconducting order parameter. 
In the discussion of our model, we demonstrated that the response is 
determined by the interlayer phase structure 
of the order parameters, but is independent of the ratio of the magnitudes of the
singlet and triplet components. 

A special situation appears for dominant spin triplet pairing, which 
has the same phase for all layers. In this case,
the spin susceptibility along the $c$-axis is the same as that in
the normal state and is basically independent of spin-orbit coupling. 
On the other hand, for the dominant spin singlet pairing, the Pauli
contribution to the spin susceptibility is completely suppressed at $T=0$.
Here, only the Van-Vleck susceptibility induced by layer-modulated
Rashba spin-orbit coupling yields a finite contribution, depending on
the interlayer coupling. 
The detailed orbital symmetry of order parameters, such as the $s$-, $p$-, 
and $d$-wave, affects these findings only weakly. 

Our study gives the first account of the spin polarizability of superconductivity in
superconductors with local non-centrosymmetricity in clean artificially layered 
superconductors. Previous studies based on similar concepts have addressed 
dirty $s$-wave superconductors~\cite{Adams}
and random spin triplet superconductors~\cite{Yanase_randomtriplet}. 
Thus, it is also interesting to review superconductors with intrinsic multilayer 
structures, such as some high-$T_{\rm c}$ cuprates, focusing on the 
role of broken local inversion symmetry. An additional feature that has been
discussed recently is the effect of staggered antisymmetric spin-orbit coupling
on superconductivity in some classes of centrosymmetric crystals that also
belong to a similar class \cite{Fischer-2011}. 

Finally, we return to the system that initially triggered our study, the superlattice 
of CeCoIn$_5$~\cite{private.superlattices}. 
In view of the fact that bulk CeCoIn$_5$ is known to realize 
spin singlet superconductivity, we believe that the dominantly 
spin singlet pairing state is relevant for the multilayer CeCoIn$_5$ 
in the superlattice. 
Unfortunately, experimental data of spin susceptibility in the 
superconducting state are not yet available. However, the effect of 
superconductivity on the spin susceptibility may be roughly estimated 
using the upper critical field $H_{\rm c2}$, because the $H_{\rm c2}$ 
of CeCoIn$_5$ is determined by the paramagnetic depairing effect. 
From a rough estimation, 
\begin{eqnarray}
H_{\rm c2}=\frac{H^{\rm P}}{\sqrt{1-\chi_{\rm s}/\chi_{\rm n}}},
\label{ucf}
\end{eqnarray} 
where $H^{\rm P}$ is the Pauli-limited upper critical field 
in conventional spin singlet superconductors,  
the large enhancement of the upper critical fields observed in the superlattice of 
CeCoIn$_5$  might be caused by 
spatially modulated Rashba spin-orbit coupling and may not necessary be a signature of
strong-coupling effects \cite{private.superlattices}. 
Moreover, we believe that the variability of the superlattices and also the possibility 
of the local measurements of magnetic properties by NMR measurement would give new insights 
into spin-orbit coupling in these artificial systems.
Moreover, in this context, the fate of the magnetic quantum critical point of CeCoIn$_5$ is
an issue that arouses experimental and theoretical interest.

\section*{Acknowledgements}
The authors are grateful to H. Shishido, T. Shibauchi, Y. Matsuda, 
M. Fischer, and D. F. Agterberg
for fruitful discussions. 
D. M. thanks Y. Yamakawa for help with the numerical calculation. 
This work was supported by a Grant-in-Aid for Scientific Research 
on Innovative Areas ``Heavy Electrons'' (No. 21102506) from MEXT, 
Japan. It was also supported by a Grant-in-Aid for Young Scientists 
(B) (No. 20740187) from JSPS. 
We are also grateful for the financial support from the Swiss Nationalfonds, the NCCR MaNEP, 
and the Pauli Center of ETH Zurich.


\begin{thebibliography}{10}

\bibitem{PhysRevLett.92.027003}
E.~Bauer, G.~Hilscher, H.~Michor, C.~Paul, E.~W. Scheidt, A.~Gribanov,
  Y.~Seropegin, H.~No\"el, M.~Sigrist, and P.~Rogl: Phys. Rev. Lett. {\bfseries
  92} (2004) 027003.

\bibitem{Springer}
To be published in {\it Non-centrosymmetric Superconductivity}, ed. M. Sigrist and E. Bauer (Springer). 

\bibitem{LowTempPhys.31.748}
E.~Bauer, I.~Bonalde, and M.~Sigrist: Low. Temp. Phys. {\bfseries 31} (2005)
  748.

\bibitem{JPSJ.76.051009}
E.~Bauer, H.~Kaldarar, A.~Prokofiev, E.~Royanian, A.~Amato, J.~Sereni,
  W.~Br\"{a}mer-Escamilla, and I.~Bonalde: J. Phys. Soc. Jpn. {\bfseries 76} (2007)
  051009.

\bibitem{JPSJ.73.3129}
T.~Akazawa, H.~Hidaka, H.~Kotegawa, T.~C. Kobayashi, T.~Fujiwara, E.~Yamamoto,
  Y.~Haga, R.~Settai, and Y.~\={O}nuki: J. Phys. Soc. Jpn {\bfseries 73} (2004)
  3129.

\bibitem{PhysRevLett.95.247004}
N.~Kimura, K.~Ito, K.~Saitoh, Y.~Umeda, H.~Aoki, and T.~Terashima: Phys. Rev.
  Lett. {\bfseries 95} (2005) 247004.

\bibitem{JPSJ.76.051010}
N.~Kimura, Y.~Muro, and H.~Aoki: J. Phys. Soc. Jpn {\bfseries 76} (2007)
  051010.

\bibitem{JPSJ.75.043703}
I.~Sugitani, Y.~Okuda, H.~Shishido, T.~Yamada, A.~Thamizhavel, E.~Yamamoto,
  T.~D. Matsuda, Y.~Haga, T.~Takeuchi, R.~Settai, and Y.~\={O}nuki: J. Phys.
  Soc. Jpn {\bfseries 75} (2006) 043703.

\bibitem{JPSJ.76.051003}
R.~Settai, T.~Takeuchi, and Y.~\={O}nuki: J. Phys. Soc. Jpn {\bfseries 76}
  (2007) 051003.
 
\bibitem{rf:CeCoGe} 
M. Measson, R. Settai, and Y. \={O}nuki: private communication; 
see also, A. Thamizhavel, H. Shishido, Y. Okuda, H. Harima, T. D. Matsuda, 
Y. Haga, R. Settai, and Y. \={O}nuki: J. Phys. Soc. Jpn. {\bf 75} (2006) 044711. 

\bibitem{rf:togano}
K. Togano, P. Badica, Y. Nakamori, S. Orimo, H. Takeya, and K. Hirata:
Phys. Rev. Lett. {\bf 93} (2004) 247004; 
P. Badica, T. Kondo, and K. Togano:
J. Phys. Soc. Jpn. {\bf 74} (2005) 1014; 
see also, H. Q. Yuan, D. F. Agterberg, N. Hayashi, P. Badica, 
D. Vandervelde, K. Togano, M. Sigrist, and M. B. Salamon: 
Phys. Rev. Lett. {\bf 97} (2006) 017006; 
M. Nishiyama, Y. Inada, and G.-q. Zheng: 
Phys. Rev. Lett. {\bf 98} (2007) 047002. 

\bibitem{rf:akimitsu}
C. Krupka, A. L. Giorgi, N. H. Krikorian, and E. G. Szklarz: 
J. Less-Common Met. {\bf 19} (1969) 113; 
G. Amano, S. Akutagawa, T. Muranaka, Y. Zenitani, and J. Akimitsu: 
J. Phys. Soc. Jpn. {\bf 73} (2004) 530. 

\bibitem{rf:shibayama}
T. Shibayama, M. Nohara, H. Aruga Katori, Y. Okamoto, 
Z. Hiroi, and H. Takagi: 
J. Phys. Soc. Jpn. {\bf 76} (2007) 073708. 

\bibitem{rf:klimczuk}
T. Klimczuk, Q. Xu, E. Morosan, J. D. Thompson, H. W. Zandbergen, and R. J. Cava:
\PRB \textbf{74} (2006) 220502(R); 
T. Klimczuk, F. Ronning, V. Sidorov, R. J. Cava, and J. D. Thompson: 
\PRL \textbf{99} (2007) 257004. 

\bibitem{rf:mu}
G. Mu, Y. Wang, L. Shan, and H.-H. Wen: \PRB {\bf 76} (2007) 064527. 

\bibitem{rf:zuev}
Y. L. Zuev, V. A. Kuznetsova, R. Prozorov, M. D. Vannette, M. V. Lobanov, 
D. K. Christen, and J. R. Thompson: \PRB {\bf 76} (2007) 132508. 

\bibitem{rf:kreiner}
Z. Ren, J. Kato, T. Muranaka, J. Akimitsu, M. Kriener, and Y. Maeno: 
\JPSJ {\bf 76} (2007) 103710. 


\bibitem{Sato-Fujimoto}
M. Sato, Y. Takahashi, and S. Fujimoto: Phys. Rev. Lett. {\bf 103} (2009) 020401.

\bibitem{Iskin}
M. Iskin and 
A. L. Suba\ifmmode \mbox{\c{s}}\else \c{s}\fi{}\ifmmode \imath \else \i \fi{}: 
Phys. Rev. Lett. {\bf 107} (2011) 050402. 

\bibitem{Sov.Phys.68.1244}
V.~M. Edelstein: Sov. Phys. JETP {\bfseries 68} (1989) 1244.

\bibitem{PhysRevB.72.172501}
V.~M. Edelstein: Phys. Rev. B {\bfseries 72} (2005) 172501.

\bibitem{PhysRevB.65.144508}
S.~K. Yip: Phys. Rev. B {\bfseries 65} (2002) 144508.

\bibitem{PhysRevB.72.024515}
S.~Fujimoto: Phys. Rev. B {\bfseries 72} (2005) 024515.

\bibitem{JPSJ.76.034712}
S.~Fujimoto: J. Phys. Soc. Jpn. {\bfseries 76} (2007) 034712.

\bibitem{Sov.Phys.44.1243}
L.~N. Bulaevskii, A.~A. Guseinov, and A.~I. Rusinov: Sov. Phys. JETP {\bfseries
  44} (1976) 1243.

\bibitem{PhysRevLett.87.037004}
L.~P. Gor'kov and E.~I. Rashba: Phys. Rev. Lett. {\bfseries 87} (2001) 037004.

\bibitem{PhysRevLett.92.097001}
P.~A. Frigeri, D.~F. Agterberg, A.~Koga, and M.~Sigrist: Phys. Rev. Lett.
  {\bfseries 92} (2004) 097001.

\bibitem{NewJPhys.6.115}
P.~A. Frigeri, D.~F. Agterberg, and M.~Sigrist: New J. Phys. {\bfseries 6}
  (2004) 115.

\bibitem{PhysRevLett.94.027004}
K.~V. Samokhin: Phys. Rev. Lett. {\bfseries 94} (2005) 027004.

\bibitem{PhysRevB.72.212504}
V.~P. Mineev and K.~V. Samokhin: Phys. Rev. B {\bfseries 72} (2005) 212504.

\bibitem{JPSJ.76.043712}
Y.~Yanase and M.~Sigrist: J. Phys. Soc. Jpn. {\bfseries 76} (2007) 043712.

\bibitem{JPSJ.77.124711}
Y.~Yanase and M.~Sigrist: J. Phys. Soc. Jpn. {\bfseries 77} (2008) 124711.

\bibitem{JPSJ.76.124709}
Y.~Yanase and M.~Sigrist: J. Phys. Soc. Jpn. {\bfseries 76} (2007) 124709.

\bibitem{JETP.Lett.78.637}
O.~V. Dimitrova and M.~V. Feigel'man: JETP Lett. {\bfseries 78} (2003) 637.

\bibitem{PhysRevB.76.014522}
O.~Dimitrova and M.~V. Feigel'man: Phys. Rev. B {\bfseries 76} (2007)
  014522.

\bibitem{PhysRevB.70.104521}
K.~V. Samokhin: Phys. Rev. B {\bfseries 70} (2004) 104521.

\bibitem{PhysRevLett.94.137002}
R.~P. Kaur, D.~F. Agterberg, and M.~Sigrist: Phys. Rev. Lett. {\bfseries 94}
  (2005) 137002.

\bibitem{PhysRevB.75.064511}
D.~F. Agterberg and R.~P. Kaur: Phys. Rev. B {\bfseries 75} (2007) 064511.

\bibitem{Matsunaga-Ikeda}
Y. Matsunaga, N. Hiasa, and R. Ikeda: 
Phys. Rev. B {\bf 78} (2008) 220508. 
 
\bibitem{Yanase_randomtriplet}
Y. Yanase: 
J. Phys. Soc. Jpn. {\bf 79} (2010) 084701. 

\bibitem{Science.327.980}
H.~Shishido, T.~Shibauchi, K.~Yasu, T.~Kato, H.~Kontani, T.~Terashima, and
  Y.~Matsuda: Science {\bfseries 327} (2010) 980.

\bibitem{private.superlattices}
Y. Mizukami, H.~Shishido, T.~Shibauchi, M. Shimozawa, S. Yasumoto, 
D. Watanabe, M. Yamashita, H. Ikeda, T. Terashima, H. Kontani, 
and Y.~Matsuda: Nat. Phys. {\bf 7} (2011) 849. 

\bibitem{Tayama}
T. Tayama, A. Harita, T. Sakakibara, Y. Haga, H. Shishido, R. Settai, 
and Y. Onuki: Phys. Rev. B {\bf 65} (2002) 180504. 

\bibitem{radovan2003}
H.~A. Radovan, N.~A. Fortune, T.~P. Murphy, S.~T. Hannahs, 
E.~C. Palm, S.~W. Tozer, and D. Hall:  Nature {\bf 425} (2003) 51.
\bibitem{PhysRevLett.91.187004}
A. Bianchi, R. Movshovich, C. Capan, P.~G. Pagliuso, and J.~L. Sarrao:  
Phys. Rev. Lett. {\bf 91} (2003) 187004.

\bibitem{young2007}
B.-L. Young, R.~R. Urbano, N.~J. Curro, J.~D. Thompson, 
J.~L. Sarrao, A.~B. Vorontsov, and M.~J. Graf:  
Phys. Rev. Lett. {\bf 98} (2007) 036402.
\bibitem{kenzelmann2008}
M. Kenzelmann, T. Str\"{a}ssle, C. Niedermayer, M. Sigrist, 
B. Padmanabhan, M. Zolliker, A.~D. Bianchi, R. Movshovich, 
E.~D. Bauer, J.~L. Sarrao, and J.~D. Thompson:  
Science {\bf 321} (2008) 1652.

\bibitem{paglione2003}
J. Paglione, M. A. Tanatar, D. G. Hawthorn, E. Boaknin, R. W. Hill, 
F. Ronning, M. Sutherland, L. Taillefer, C. Petrovic, and
P. C. Canfield: 
Phys. Rev. Lett. {\bf 91} (2003) 246405. 

\bibitem{bianchi2003}
A. Bianchi, R. Movshovich, I. Vekhter, P.~G. Pagliuso, and J.~L. Sarrao:  
Phys. Rev. Lett. {\bf 91} (2003) 257001.

\bibitem{ronning2005}
F. Ronning, C. Capan, A. Bianchi, R. Movshovich, A. Lacerda,
M.~F. Hundley, J.~D. Thompson, P.~G. Pagliuso, and J.~L. Sarrao:  
Phys. Rev. B {\bf 71} (2005) 104528.

\bibitem{izawa2007}
K. Izawa, K. Behnia, Y. Matsuda, H. Shishido, R. Settai, Y. Onuki, 
and J. Flouquet: Phys. Rev. Lett. {\bf 99} (2007) 147005. 

\bibitem{panarin2009}
J. Panarin, S. Raymond, G. Lapertot, and J. Flouquet: 
J. Phys. Soc. Jpn. {\bf 78} (2009) 113706. 

\bibitem{Mukuda}
H. Mukuda, M. Abe, Y. Araki, Y. Kitaoka, K. Tokiwa, T. Watanabe, 
A. Iyo, H. Kito, and Y. Tanaka: Phys. Rev. Lett. {\bf 96} (2006)
087001. 

\bibitem{Shimizu-Mukuda}
S. Shimizu, S. Tabata, H. Mukuda, Y. Kitaoka, P. M. Shirage, 
H. Kito, and A. Iyo: J. Phys. Soc. Jpn. {\bf 80} (2011) 043706. 

\bibitem{PhysRevLett.94.137003}
M.~Mori and S.~Maekawa: Phys. Rev. Lett. {\bfseries 94} (2005) 137003.

\bibitem{Maruyama_LT} 
D. Maruyama, M. Sigrist, and Y. Yanase: to be published in J. Phys.: Conf. Ser. 

\bibitem{Adams}
X. S. Wu, P. W. Adams, Y. Yang, and R. L. McCarley: 
Phys. Rev. Lett. {\bf 96} (2006) 127002. 

\bibitem{Fischer-2011}
M. H. Fischer, F. Loder, and M. Sigrist: 
Phys. Rev. B {\bf 84} (2011) 184533.

\end{thebibliography}
\end{document}